\documentclass[prl,twocolumn,aps,showpacs]{revtex4}
\usepackage{epsfig}
\usepackage{bm}
\begin{document}
\title{Competition between final-state and pairing-gap effects
              in the radio-frequency spectra of ultracold Fermi atoms}
\author{A. Perali, P. Pieri, and G.C. Strinati}
\affiliation{Dipartimento di Fisica, 
Universit\`{a} di Camerino, I-62032 Camerino, Italy}
\date{\today}

\begin{abstract}
The radio-frequency spectra of ultracold Fermi atoms are calculated by including final-state interactions affecting the excited level of the transition, and compared 
with the experimental data.
A competition is revealed between pairing-gap effects which tend to push the oscillator strength toward high frequencies away from threshold, and final-state 
effects which tend instead to pull the oscillator strength toward threshold.
As a result of this competition, the position of the peak of the spectra cannot be simply related to the value of the pairing gap, whose extraction thus 
requires support from theoretical calculations.
\end{abstract}

\pacs{03.75.Ss,03.75.Hh,32.30.Bv,74.20.-z}
\maketitle

One key issue concerning ultracold trapped Fermi atoms with a mutual attractive interaction is the detection of their superfluid behavior below a critical temperature $T_{c}$.
 As conventional techniques (like the Meissner effect) to identify the occurrence of the superconducting phase do not work for these neutral systems, additional 
physical effects need be explored to the purpose.

In this context, radio-frequency (RF) spectra were first proposed \cite{Grimm-2004} to signal the presence of a pairing gap for ultracold Fermi ($^{6}$Li) atoms at low enough 
temperature.
Collecting data at various couplings and for temperatures from below to above $T_{c}$, the evolution of these spectra was qualitatively interpreted in the 
light of an underlying pairing gap.
These experiments have since been extended to different systems \cite{Jin-2005} as well as in the presence of spin population imbalance \cite{Ketterle-I-2007}, and quite
recently also by attaining spatial resolution \cite{Ketterle-II-2007}.

Several theoretical attempts have been put forward to interpret the RF spectra of ultracold Fermi atoms 
\cite{Torma-2004,Chin-Julienne-2005,Ohashi-Griffin-2005,Levin-2005,Baym-2006,Zwerger-2007,Baym-2007}.
There is at present wide consensus \cite{GPS-RMP-2007}, however,  that this difficult problem is still unsolved theoretically, and a challenging call has been 
launched \cite{Ketterle-II-2007} for a comprehensive theoretical interpretation of RF spectra of Fermi gases in the strong-interacting regime.

A definitely important step forward for interpreting these spectra is the inclusion of \emph{final-state effects}, whereby the atom excited to the final state $|3\rangle$ 
from the initial state $|2\rangle$ interacts with the atom in the state $|1\rangle$.
Here, the states $|1\rangle$ and $|2\rangle$ participate to the Cooper pairing as modulated by the Fano-Feshbach resonance with scattering length $a_{12}$, while the (empty) 
final state $|3\rangle$ excited by the RF transition interacts with the state $|1\rangle$ via a different Fano-Feshbach resonance with scattering length 
$a_{13}$~\cite{Chin-Julienne-2005}.
This final-state interaction should considerably affect the density of states available to the excited atom in channel 1$-$3, thereby triggering a competition with 
pairing-gap effects present in channel 1$-$2.

Purpose of the present paper is to show that considering this competition considerably improves the description of the RF spectra, by accounting for 
important features that cannot be explained otherwise.
We limit here to low temperatures where the pairing gap is well developed, but extension proves feasible also above $T_{c}$ where the pseudogap takes 
the place of the pairing gap.

On physical grounds, the approach discussed in the present paper bears strong analogies with the treatment of the exciton effect in the optical spectra of semiconductors 
\cite{Hanke-Sham-1975}.
In that case, the long-range Coulomb attraction considerably modifies the absorption spectrum via the density of final states, leading to a finite 
(in contrast to a vanishing) value of the absorption coefficient at threshold \cite{Bassani-1975}. 
For the RF spectra of ultracold Fermi atoms, final-state effects are less dramatic owing to the short-range nature of the interactions in both 1$-$2 and 1$-$3 
channels (which are assimilated by attractive delta-functions with scattering lengths $a_{12}$ and $a_{13}$, respectively).
Yet, we shall see that for large enough $|a_{13}|$ these exciton-like effects modify the RF spectra in a definitive fashion.

The importance of taking the effect of $a_{13}$ into account was demonstrated in Ref.~\cite{Chin-Julienne-2005} for the molecular (two-body) case, which represents the 
extreme BEC limit of the BCS-BEC crossover.
In addition, Ref.~\cite{Baym-2006} considered the effect of $a_{13}$ on the $f$-sum rule within the BCS-RPA approximation, albeit only in 
the BCS limit of the BCS-BEC crossover.
In the present paper, we adopt a \emph{diagrammatic representation} of the correlation function relevant to the calculation of the RF spectra of ultracold Fermi atoms, to 
include final-state on top of pairing-gap effects.
This representation recovers the results of Refs.~\cite{Chin-Julienne-2005} and \cite{Baym-2006} in the respective limits.

Besides the exciton-like effect between the photo-excited atom and the Cooper-pair-mate left behind, t-matrix self-energy corrections of level $|3\rangle$ due to interaction 
with level $|1\rangle$ may account for additional final-state effects associated with the action of the medium on the photo-excited atom. These self-energy corrections, 
however, should not be included if atoms in level $|3\rangle$ do not reach thermal equilibrium with the environment in the time scale of the experiment.  
In addition, pairing fluctuations beyond mean field~\cite{PPS-2004} are needed for a fuller description of the initial state of the RF transition.

Self-energy corrections of level $|3\rangle$ due to interaction with level $|2\rangle$ are instead dropped altogether, because they are irrelevant in the molecular  
limit and the interaction 2$-$3 does not affect the value of the $f$-sum rule whenever level $|3\rangle$ is empty~\cite{Baym-2006}. 
As the presence of a t-matrix self-energy in channel 2$-$3 would spoil this property, inclusion of additional diagrams for the response function would 
be required to recover it. 

In RF experiments with $^{6}$Li atoms, a rf pulse with frequency $\omega_{L}$ transfers population from state $|2\rangle$ to state 
$|3\rangle$ and the atom loss from state $|2\rangle$ is measured. 
For weak excitations, the RF signal $\delta \langle I(\omega)\rangle$ can be calculated within linear-response theory yielding \cite{Ohashi-Griffin-2005}:
\begin{equation}
\delta \langle I(\omega)\rangle \, = \, - \, 2 \, \gamma^{2} \, \int \! \, d\mathbf{r} \,
d\mathbf{r'} \, \, \mathrm{Im} \{\Pi^{R}(\mathbf{r},\mathbf{r'};\omega)\} \,\,\, 
                                                                            \label{current-RF}
\end{equation}
where $\gamma$ contains the atomic matrix element of the transition and $\Pi^{R}(\mathbf{r},\mathbf{r'};\omega)$ is the Fourier transform of the retarded correlation function 
$\Pi^{R}(\mathbf{r},\mathbf{r'};t-t') = -i\theta(t-t') \langle [B(\mathbf{r},t),B(\mathbf{r'},t')]\rangle$ taken at the frequency 
$\omega=\omega_{L}+ \mu-\mu_{3}$.
Here, $B^{\dagger}(\mathbf{r})=\psi_{3}^{\dagger}(\mathbf{r}) 
\psi_{2}(\mathbf{r})$ is the relevant local transition operator and $\langle\cdots\rangle$  a thermal average, while $\mu$ and $\mu_{3}$ are the chemical potentials for the 
initial (populated) and final (empty) levels, in the order. 

Expression (\ref{current-RF}) applies to alternative dynamical approximations. Following a standard approach for quantum 
many-body systems \cite{FW}, the retarded correlation function is calculated via its Matsubara counterpart
\begin{eqnarray}
\Pi(\mathbf{r},\mathbf{r'};\Omega_{\nu}) &=& \int_{0}^{\beta} \! 
 d\tau \, e^{i \Omega_{\nu} \tau}
\langle\mathrm{T}_{\tau}[\psi_{2}(\mathbf{r'},0)\psi_{2}^{\dagger}(\mathbf{r},\tau^{+})\nonumber\\ 
&\times&\psi_{3}(\mathbf{r},\tau)\psi_{3}^{\dagger}(\mathbf{r'},0^{+})] \rangle
\label{Matsubara-correlation-function}
\end{eqnarray}
where $\mathrm{T}_{\tau}$ is the time-ordering operator for imaginary time $\tau$ and $\Omega_{\nu}=2\pi\nu/\beta$ ($\nu$ integer) is a bosonic Matsubara frequency with 
inverse temperature $\beta$.
Analytic continuation from $i\Omega_{\nu}$ to $\omega+i\eta$ (where $\eta$ is a positive infinitesimal) is taken eventually.

Previous calculations \cite{Torma-2004,Ohashi-Griffin-2005,Levin-2005} attempting comparison with experiments considered a mean-field factorization of the thermal 
average in Eq.(\ref{Matsubara-correlation-function}), resulting in the approximate expression
\begin{equation}
\Pi(\mathbf{r},\mathbf{r'};\Omega_{\nu}) \simeq  \frac{1}{\beta} 
\sum_{n}  G_{2}(\mathbf{r'},\mathbf{r};\omega_{n}) 
G_{3}(\mathbf{r},\mathbf{r'};\omega_{n}+\Omega_{\nu})
\label{BCS-bubble}
\end{equation}
where $\omega_{n}=(2n+1)\pi/\beta$ ($n$ integer) is a fermionic Matsubara frequency and the $G$'s are single-particle propagators for the initial $|2\rangle$ and final 
$|3\rangle$ levels of the transition.
This expression is represented by the ``bubble'' of
Fig.~1(a).
The RF spectrum calculated within the approximation (\ref{BCS-bubble}) shows the following features: 
(i) In the absence of interactions, the spectrum reduces to a Dirac delta-function $2 \pi \gamma^{2} N_{2} \delta(\omega_{L}-\omega_{a})$ where $\omega_{a}$ is the frequency
of the atomic transition and $N_{2}$ the total number of atoms originally in state $|2\rangle$; 
(ii) Pairing interactions between states $|1\rangle$ and $|2\rangle$ shift this delta peak to higher frequencies and broaden it to the extent that it behaves like 
$\omega^{-3/2}$ for frequencies larger than the relevant fermionic energy scales, while preserving the \emph{sum rule} for the total area of the spectrum \cite{footnote-1}
\begin{equation}
\int _{- \infty}^{+ \infty} \!  d\omega \, \delta \langle I(\omega)\rangle=2\pi\gamma^{2} N_{2}                               
\label{sum-rule}
\end{equation}
(we assume throughout that level $|3\rangle$ is empty).

% Figure 1
\begin{figure}
\begin{center}
\epsfxsize=7.5cm
\epsfbox{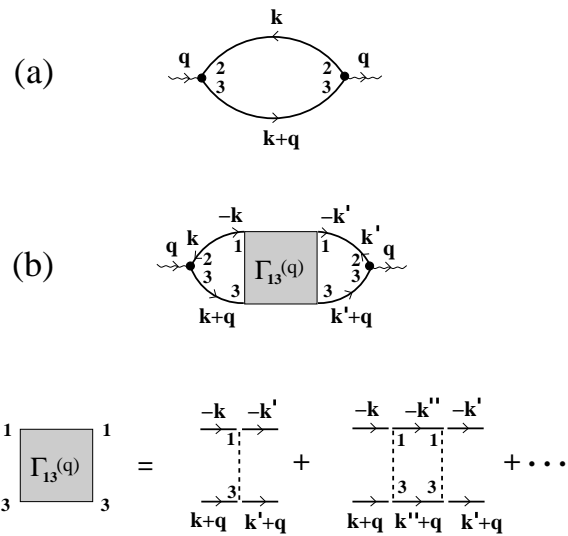}
\caption{Diagrammatic representation of the correlation function (\ref{Matsubara-correlation-function}): 
(a) BCS bubble and (b) BCS-RPA series.
Full and broken lines stand for fermionic single-particle propagators and interactions, respectively, while numbers attached to the end points identify the relevant states.
The wiggly lines represent the coupling to the RF field. For simplicity, all diagrams are drawn in four-momentum space.} 
\end{center}
\end{figure}

The slow convergence for large $\omega$ resulting from the approximation (\ref{BCS-bubble}) appears evident when one attempts comparison with the experimental spectra 
\cite{Torma-2004,Ohashi-Griffin-2005,Levin-2005}. 
It further leads to a divergent value for the $f$-sum rule that corresponds to the first moment of the spectrum $\delta \langle I(\omega)\rangle$.
In this respect, it was pointed out in Ref.~\cite{Baym-2006} that a finite value for the $f$-sum rule results instead by going beyond the bubble approximation 
(\ref{BCS-bubble}) for $\delta \langle I(\omega) \rangle$ and summing the BCS-RPA series, in analogy to the theory of superconductivity when implementing 
gauge invariance for the response functions \cite{Schrieffer}.
In Ref.~\cite{Baym-2006} this extension was regarded appropriate to the weakly-interacting (BCS) regime of the BCS-BEC crossover, where comparison with the experimental 
spectra is hardly possible.

Figure 1(b) shows the diagrammatic representation of the BCS-RPA series \cite{footnote-2} contributing to the correlation function (\ref{Matsubara-correlation-function}), in 
terms of which one can gain insights about the inclusion of the physical effects discussed above, namely:
(i) The occurrence of pairing in channel 1-2 (the pairing-gap effect), which can be dealt with even by including fluctuations beyond BCS mean-field; 
(ii) The presence of an attractive interaction in channel 1-3 (the exciton-like effect), which is evident in the ladder $\Gamma_{{\rm 13}}$; 
(iii) The possible inclusion of the t-matrix self-energy dressing level $|3\rangle$ (the medium effect).
[Pairing fluctuations and t-matrix effects need not be included in the internal lines defining $\Gamma_{{\rm 13}}$.]
Note that both pairing-gap and exciton-like effects are also borne out in the two-body calculation of Ref.~\cite{Chin-Julienne-2005}. 

We have indeed verified that the results of the two-body calculation of Ref.~\cite{Chin-Julienne-2005} can be recovered from the many-body calculation of 
$\delta\langle I(\omega)\rangle$ based on the diagrams of Figs.~1(a) and 1(b), when approaching the BEC limit 
$1\ll (k_{F} a_{12})^{-1}$ where $k_{F}$ is the Fermi wave vector associated with $N_{2}$.
By this remark, we can rely on the diagrams of Figs.~1(a) and 1(b) to describe the RF spectrum $\delta \langle I(\omega)\rangle$ throughout the whole BCS-BEC crossover, the 
only limitation being the condition of low-enough temperatures such that the pairing gap is well developed.

The exciton-like effect contained in the diagrams of Fig.~1(b) generates a strong suppression of the tail of RF spectra, by augmenting its power-law 
convergence by one unity to $\omega^{-5/2}$.
In addition, the diagrams of Fig.~1(b) give a vanishing contribution to the sum rule (\ref{sum-rule}) (which, otherwise, could not be 
preserved).
This implies that, to the negative tail for large $\omega$ resulting from the diagrams of Fig.~1(b), there corresponds a positive pile up for small $\omega$ which tends to 
peak the spectrum toward threshold.

% Figure 2
\begin{figure}
\begin{center}
\epsfxsize=5cm
\epsfbox{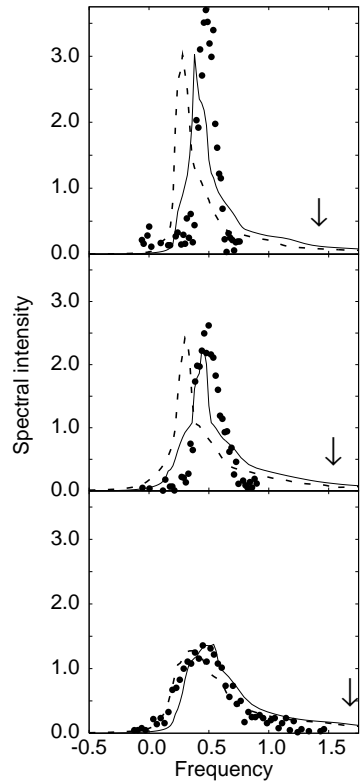}
\caption{Comparison between theoretical [without (full lines) and with (broken lines) inclusion of the self-energy of level $|3\rangle$] and experimental (dots) RF spectra for $(k_{F} a_{12})^{-1}=0$ and
$(k_{F} a_{13})^{-1}=$ $-0.93$ (upper panel),  $-1.03$ (middle panel),
and $-1.32$ (lower panel).
These panels correspond to Figs.~2(b), 2(c), and 2(d) of Ref.~\cite{Ketterle-II-2007}, in the order, with radial position $r=0, 0.36$, and $0.59$ in units of the Thomas-Fermi 
radius, and associated temperatures $0.042$, $0.052$, and $0.085$ in units of the local Fermi temperature.
The value $a_{13}= -174.6$ nm is taken from Ref.~\cite{Bartenstein}, while the local Fermi wave vector is taken from Ref.~\cite{Ketterle-II-2007}.
The frequency is in units of the local Fermi energy and its zero is taken at the atomic transition. 
The spectral intensity is normalized so that the right-hand side of the sum-rule (4) is unity, while 
the arbitrary scale of the experimental data is fixed to match the theoretical peak height in the lower panel.} 
\end{center}
\end{figure}

Figure 2 presents the results of our numerical calculation that includes pairing-gap (with fluctuations), exciton-like, and medium effects at the unitary limit 
$(k_{F} a_{12})^{-1}=0$, and compares them with the tomographic spectra of Ref.~\cite{Ketterle-II-2007}.
Good agreement between the theoretical and experimental spectra results from this comparison especially when the self-energy of level $|3\rangle$ is not included, thus 
suggesting that atoms in level $|3\rangle$ do not thermalize in the course of the experiment.
In particular, our calculation is able to capture the main features of the experimental spectra, namely:
(i) The value of the peak position normalized to the local Fermi energy, which increases only slightly from the trap center to the edge;
(ii) Their asymmetry, the line shape being sharper on the low-frequency side of the peak;
(iii) The strong progressive broadening of the spectra away from the trap center, which is accompanied by a lowering of the peak height.

Note that the overall profiles of Fig.~2 reveal at glance the importance of the scattering length $a_{13}$ for the RF spectra, since at unitarity for $a_{12}$ spectra 
corresponding to different local densities present a non-universal behavior that can only originate from the presence of an additional length scale in the problem.
Note further that the first moment of these spectra (identified by an arrow in each panel) does not relate to the position of the experimental peak, being moved to higher 
frequencies owing to the prominent role of the tail of the spectrum.

% Figure 3
\begin{figure}
\begin{center}
\epsfxsize=8cm
\epsfbox{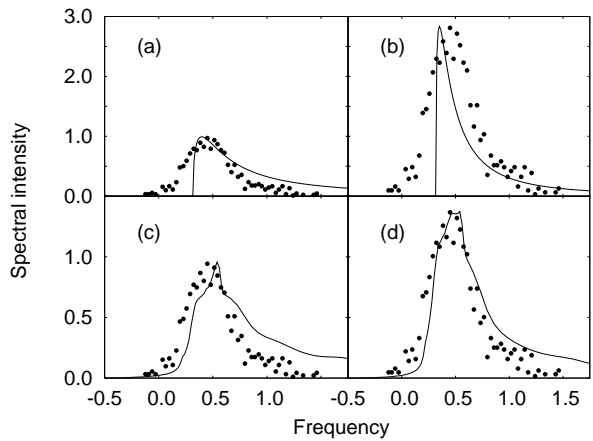}
\caption{Evolution of the theoretical spectra with the progressive refinement of the approximations. In each plot, the experimental data 
have been normalized to match the theoretical peak height. (See text for details.)} 
\end{center}
\end{figure}

To appreciate how the different physical effects contribute to the theoretical spectra of Fig.~2, we show in Fig.~3 how these spectra evolve with the 
progressive refinements of 
the approximations (for definiteness, we show the results for the lower panel of Fig.~2).
We begin in Fig.~3(a) with the BCS bubble of Fig.~1(a), evaluated with the BCS gap and chemical potential.
We further include in Fig.~3(b) the BCS-RPA series of Fig.~1(b), with the same BCS values for the gap and chemical potential.
We return in Fig.~3(c) to the bubble of Fig.~1(a), where we now include pairing fluctuations beyond mean field \cite{PPS-2004} in the fermionic propagator 
$G_2$.
Finally, in Fig.~3(d) we report the full calculation with all diagrams of 
Fig.~1 included, and with pairing fluctuations consistently included in all fermionic propagators whose level $|2\rangle$ couples to the external field
(while no self-energy correction is considered for the fermionic propagator $G_3$).
In both Figs.~3(c) and 3(d) the gap and chemical potential are evaluated, too, with the inclusion of pairing fluctuations.
In all cases, numerical accuracy results in the sum rule (\ref{sum-rule}) being satisfied at worse within 
$5\%$.

The competition between pairing-gap and final-state effects on the RF spectra can be traced by examining in detail the above evolution. 
The simplest approximation of the BCS bubble shown in Fig.~3(a) contains pairing-gap effects, which shift the peak of the spectrum to higher frequencies with respect to the 
atomic transition, but does not include any final-state effect.
The exciton-like effect is evident in Fig.~3(b), whereby the oscillator strength is pulled toward threshold with respect to Fig.~3(a) and the tail is strongly suppressed 
correspondingly.
The role of the pairing gap is further emphasized by the inclusion of pairing fluctuations beyond mean field in Fig.~3(c), which give rise to additional pseudogap effects 
that suppress the density of states near threshold and broaden the low-frequency side of the peak.
The competing role of the exciton-like effect is again operative in Fig.~3(d), resulting in the RF spectrum that matches best the experimental data.

We can finally compare the value of the peak position of the calculated RF spectra with the value of the pairing gap for a 
homogeneous system at unitarity.
For this comparison we consider the full line in the upper panel of Fig.~2 at the lowest temperature, from which we extract the value $0.38$ for the peak position 
(normalized to the local Fermi energy), while for the  pairing gap Ref.~\cite{PPS-2004} provides the value $0.53$ (the pairing gap extracted from the single-particle 
spectral function and order parameter about coincide for these coupling and temperature).
This comparison confirms the role of the competition of the physical effects that we have discussed, leading us to conclude that, 
although RF spectroscopy undoubtedly provides a direct 
signature for the presence of a pairing gap in ultracold Fermi atoms, the numerical value of this quantity cannot be simply 
extracted from the peak position of the RF spectra.
\acknowledgments
Partial support by the Italian MUR under Contract Cofin-2005 ``Ultracold Fermi Gases and Optical Lattices'' is acknowledged.


\begin{thebibliography}{99}
\bibitem{footnote-1} The sum rule (\ref{sum-rule}) can be quite generally proved from the form (\ref{Matsubara-correlation-function}) of the correlation function, and is satisfied in particular by its bubble approximation (\ref{BCS-bubble}).
\end{thebibliography}
\end{document}